\documentclass{article}

\usepackage{amsmath, amssymb}
\usepackage{graphicx}

\def\>{\ensuremath{\rangle}}
\def\<{\ensuremath{\langle}}
\def\h{\ensuremath{\mathcal{H}}}

\def\m{\ensuremath{\mathcal{M}}}
\def\n{\ensuremath{\mathcal{N}}}
\def\ch{\h_1\otimes \cdots\otimes \h_n}
\def\i{\ensuremath{{\rm i}}}

\def\supp{\ensuremath{{\rm supp}}}
\def\span{\ensuremath{{\rm span}}}
\def\exp{\ensuremath{{\rm exp}}}
\def\ind{\ensuremath{\mathcal{I}}}
\newcommand{\define}{\equiv}

\newcommand{\tr}{{\rm Tr}}

\newtheorem{corollary}{Corollary}
\newtheorem{definition}{Definition}

\newtheorem{proposition}{Proposition}
\newtheorem{theorem}{Theorem}

\begin{document}
\title{Locally undetermined states, generalized Schmidt decomposition, and an application in distributed computing}

\author{Yuan Feng\thanks{Email: feng-y@tsinghua.edu.cn}, Runyao
Duan\thanks{Email: dry@tsinghua.edu.cn}, and Mingsheng Ying\thanks{Email: yingmsh@tsinghua.edu.cn}\\ \\
Tsinghua National Laboratory for Information Science and
Technology(TNList)\\
State Key Laboratory of Intelligent Technology and Systems,\\
Department of Computer Science and Technology,\\
Tsinghua University, Beijing, China.}
 \maketitle
\thispagestyle{empty}

\begin{abstract}
Multipartite quantum states that cannot be uniquely determined by
their reduced states of all proper subsets of the parties exhibit
some inherit `high-order' correlation. This paper elaborates this
issue by giving necessary and sufficient conditions for a pure
multipartite state to be locally undetermined, and moreover,
characterizing precisely all the pure states sharing the same set of
reduced states with it. Interestingly, local determinability of pure
states is closely related to a generalized notion of Schmidt
decomposition. Furthermore, we find that locally undetermined states
have some applications to the well-known consensus problem in
distributed computation. To be specific, given some physically
separated agents, when communication between them, either classical
or quantum, is unreliable and they are not allowed to use local
ancillary quantum systems, then there exists a totally correct and
completely fault-tolerant protocol for them to reach a consensus if
and only if they share a priori a locally undetermined quantum
state.
\end{abstract}

\section{Introduction}

Entanglement is a striking feature of quantum mechanics which plays
a central role in quantum computation and quantum information
processing tasks such as quantum teleportation, superdense coding,
and cryptographic protocols, etc \cite{NC00}. In some sense, the
advantage of quantum computation and quantum information processing
over their classical counterparts is exactly due to the existence
and proper use of entanglement. As a result, the theory of
entanglement is important both theoretically and practically, and
has been widely investigated in the past several decades.

Characterizing different types of entanglement is one of the most
active research fields in entanglement theory. For multipartite
states, one way towards such a characterization is to examine local
determinability of them: if a quantum state shared among $n$ parties
cannot be uniquely determined by its reduced states of fewer than
$n$ parties, then in a sense the state exhibits `higher-order'
entanglement which is not attributable to all `lower-order'
entanglement among these parties. Surprisingly, Linden $et$ $al$.
\cite{LPW02,LW02,JL05} showed that in pure state case, chance for
the existence of such `higher-order' entanglement is very little. To
be specific, almost all $n$-party pure states are determined by
their reduced states of less than $n$ parties. In fact, when the
number of parties is sufficiently large, for almost all states
except for a zero measure set, about two-thirds of the parties are
sufficient to determine the global pure state. At the other extreme,
Di\'{o}si \cite{Di04} presented a method to construct a generic
3-qubit pure state from its three 2-qubit reduced states.

Although the set of locally undetermined pure states is proven to be
zero measure, describing it precisely might be useful, as pointed
out by Linden and Wootters \cite{LW02}, in investigating properties
of multiparticle entanglement. Along this line, Walck and Lyons
\cite{WL08,WL082} showed that in the special case of qubit systems,
the only possible locally undetermined states are generalized
GHZ-states. The main purpose of the current paper is to extend their
result to the general case where Hilbert spaces with arbitrary
dimensions are permitted. We present necessary and sufficient
conditions for a multipartite pure state to be locally undetermined,
and when a state is locally undetermined, we give the explicit form
of all the pure states which share the same set of reduced states.
Especially, we find that local determinability of pure states is
closely related to a generalized notion of Schmidt decomposition
which, to our best knowledge, is first defined in the present paper.

Distributed consensus is one of the central problems in distributed
algorithms where a group of physically separated but
inter-communicating agents need to reach agreement \cite{Ly96}. It
has promising applications in distributed data processing and file
management. In classical case, however, no deterministic protocol
exists in an asynchronous setting which guarantees the correct
agents to reach a consensus within finitely many steps, if some
agents might fail during executing the protocol~\cite{FLP85}. Even
if probabilistic protocols are allowed, only one half of fail-stop
faulty agents or one-third of malicious agents are tolerated if the
probability of reaching agreement is required to be one \cite{BT85}.

D'Hondt and Panangaden first investigated distributed consensus with
the aid of quantum resources \cite{DP05}. They proved that GHZ
state, or GHZ-like states in higher dimensional case, is the only
possible pure states to give a totally correct solution to the
distributed consensus problem for an anonymous network in a purely
quantum way. Here a protocol is called totally correct if it
successfully terminates with its goal achieved within finitely many
steps along each computation path, and it is purely quantum if no
classical post-processing is allowed during the execution. The
striking feature of GHZ-like states as quantum resource in solving
distributed consensus is that they can not only solve the problem,
but more importantly, the solution is fault-tolerant in the sense
that no matter how unreliable the communication channels are -- even
if the communication, classical or quantum, is forbidden at all --
and how many agents fail, the correct agents can still reach a
consensus. We call this property completely fault-tolerant, which
should be compared with the notion of fault-tolerance considered in
\cite{BH05} (which is usually assumed in classical setting) where
faults are modeled by unpredictable behavior of some agents while
the message exchange between agents is perfect. In this paper, we
extend the result of D'Hondt and Panangaden by considering a more
general network, anonymous or not, where a multipartite pure state
is shared between the agents but any local ancillary quantum system
is forbidden. Interestingly, we find that a totally correct and
completely fault-tolerant protocol exists if and only if the shared
state is locally undetermined.

\section{Generalized Schmidt decomposition of multipartite pure states}

This section is devoted to the definition of a generalized Schmidt
decomposition of pure states in multipartite Hilbert space. Let
$\rho$ be a density operator and $\rho=\sum_{i=1}^M\lambda_i
|i\>\<i|$ be its spectrum decomposition. Then
$\supp(\rho)=\span\{|i\> : 1\leq i\leq M\}$. For a set of density
operators $\rho_1, \dots, \rho_n$, we define
$\supp\{\rho_1,\dots,\rho_n\}=\sum_{i=1}^n \supp(\rho_i)$. Given a
multipartite pure state $|\psi\>\in \ch$ and $1\leq k\leq n$, we
denote by $\rho^{\psi}_k$ the 1-party reduced state of $|\psi\>$ on
the $k$-th component subsystem, i.e.,
$\rho^{\psi}_k=\tr_{\bar{k}}|\psi\>\<\psi|$ where $\bar{k}$
indicates the Hilbert space $\bigotimes_{i\neq k} \h_i\define
\h_{\bar{k}}$.

\begin{figure}[t]\centering
\includegraphics{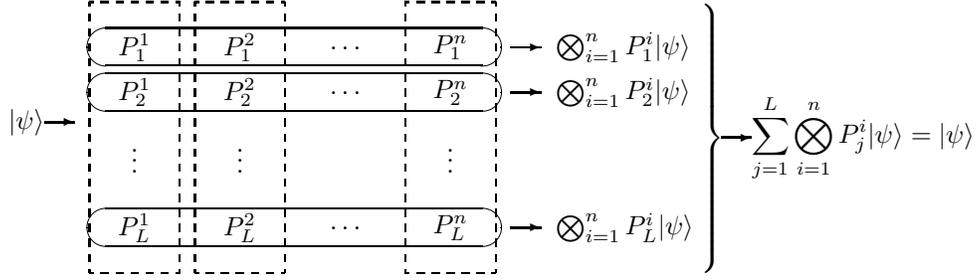}
\caption{Illustration of the Schmidt projectors of a pure state
$|\psi\>\in \ch$. Here projectors in each column constitute an
orthogonal set in the corresponding Hilbert space.} \label{fig 1 }
\end{figure}

\begin{definition}\label{def-decom} Let $|\psi\>\in \ch$. A family
$\{P_j^i : i=1,\dots,n; j=1,\dots, L\}$ of projectors are said to be
Schmidt projectors of $|\psi\>$ if
\begin{enumerate}
\item for any fixed $i$, $P_j^i : j=1,\dots, L$ are pairwise
orthogonal projectors on $\h_i$,
\item $P_j^i|\psi\>\neq 0$ for each $i$ and $j$,
\item $|\psi\>=\sum_{j=1}^L \bigotimes_{i=1}^n P_j^i|\psi\>.$
\end{enumerate}
The projectors $P_j^i$ are illustrated in Fig. \ref{fig 1 }.
\end{definition}

The Schmidt number of $|\psi\>$, $Sch(\psi)$, is defined to be the
maximal $L$ such that $|\psi\>$ has $L$ rows of Schmidt projectors,
that is
$$Sch(\psi)=\max \{L : |\psi\> \mbox{ has Schmidt projectors } P_j^i, 1\leq i\leq n, 1\leq
j\leq L\}.$$ Note that every pure state has the trivial Schmidt
projectors $\{I_{\h_i} : i=1,\dots,n\}$ where $I_{\h_i}$ is the
identity projector on $\h_i$. So the notion $Sch(\psi)$ is
well-defined. When $Sch(\psi)>1$, $|\psi\>$ is said to be
generalized Schmidt decomposable (GSD). If each Schmidt projector
can be taken to be rank 1, then $|\psi\>$ is called completely GSD.
It is well known that every bipartite pure state is completely GSD.

The next proposition indicates that we can put more constraints on
the projectors which serve as the Schmidt projectors of a pure
state. These constraints are useful in proving some later results in
this paper.

\begin{proposition}\label{prop-main} Let $|\psi\>\in \ch$. Then
$|\psi\>$ is GSD if and only if $|\psi\>$ has Schmidt operators
$\{P_j^i : i=1,\dots,n; j=1,\dots, L\}$, $L>1$,  such that for any
fixed $i$, $\{P_j^i : j=1,\dots, L\}$ constitute a projective
measurement in $\supp(\rho_i^\psi)$, that is, $P_j^i$ are pairwise
orthogonal and $\sum_{j=1}^L P_j^i = P_{\supp(\rho_i^\psi)}$. Here
for a Hilbert space $\h$, we denote by $P_{\h}$ the projector onto
$\h$.
\end{proposition}
{\it Proof.} Suppose we are given a set $\{P_j^i : i=1,\dots, n;
j=1,\dots, L\}$ where for each $i$, $P^i_j$ are pairwise orthogonal
projectors on $\h_i$, $P_j^i|\psi\>\neq 0$, and $|\psi\>=
\sum_{j=1}^L \bigotimes_{i=1}^n P_j^i |\psi\>$. Fix arbitrarily
$1\leq k\leq n$ and let $P= P_{\supp(\rho_k^\psi)}$. In the
following we prove that $\{PP_j^k P : j=1,\dots, L\}$ constitute a
projective measurement in $\supp(\rho_k^\psi)$. To this end, we need
to show: (i) $PP_j^k P$ is a nonnull projector for each $j$; (ii)
$PP_j^k P$ are pairwise orthogonal; (iii) $\sum_{j=1}^L PP_j^k P =
P$.

Take arbitrarily $1\leq j\leq L$ and $1\leq k'\leq n$ such that
$k'\neq k$. Let $P_{\bar{j}}^{k'} = \sum_{l\neq j} P_l^{k'}.$ It is
easy to observe that
$$P_j^k
|\psi\>=\bigotimes_{i=1}^n P_j^i |\psi\>,\hspace{2em}
P_{\bar{j}}^{k'} |\psi\>=\sum_{l\neq j} \bigotimes_{i=1}^n P_l^i
|\psi\>.$$ Hence
\begin{equation}\label{eq-tp-2-1}
P_j^k |\psi\>+P_{\bar{j}}^{k'} |\psi\>=|\psi\>.
\end{equation}
Let
\begin{equation}\label{eq-schdecom-7.9}
|\psi\>=\sum_{i=1}^{M_k}\sqrt{\lambda_i}|i\>_k
|\psi_i\>_{\bar{k}},\hspace{2em} \lambda_1,\dots,\lambda_{M_k}>0
\end{equation}
be the (ordinary) Schmidt decomposition of $|\psi\>$ when treated as
a bipartite state between $\h_k$ and $\h_{\bar{k}}$, where $M_k\leq
d_k$, and $\{|i\> : i=1,\dots, d_k\}$ is an orthonormal basis of
$\h_k$. Then we have from Eq.(\ref{eq-tp-2-1})
$$\sum_{i=1}^{M_k}\sqrt{\lambda_i}\big(P_j^k|i\>_k\big) |\psi_i\>_{\bar{k}} +
\sum_{i=1}^{M_k}\sqrt{\lambda_i}|i\>_k
\big(P_{\bar{j}}^{k'}|\psi_i\>_{\bar{k}}\big)=\sum_{i=1}^{M_k}\sqrt{\lambda_i}|i\>_k
|\psi_i\>_{\bar{k}}.$$ Multiplying both sides of the above equation
by $\<l|\<\psi_i|$ for any $1\leq i\leq M_k$ and $M_k< l \leq d_k$,
we have $\<l|P_j^k|i\>=0.$ That is, for each $i=1,\dots, M_k$,
$P_j^k|i\>$ lies in the Hilbert space
$\supp(\rho_k^\psi)=\span\{|i\> : 1\leq i\leq M_k\}$. As a
consequence, we have $PP_j^k|i\>=P_j^k|i\>$, and then
$PP_j^kP|\phi\>=P_j^kP|\phi\>$ for any $|\phi\>\in \h_k$. So
$PP_j^kP=P_j^kP$. On the other hand, from the fact that
$\sum_{j=1}^L P_j^k |\psi\> = |\psi\>$ and
Eq.(\ref{eq-schdecom-7.9}) we have $\sum_{j=1}^L P_j^k|i\>=|i\>$ for
each $1\leq i\leq M_k$. Thus $\sum_{j=1}^L P_j^k P=P$.

Now we can check (i), (ii), and (iii) directly. For (i), we notice
that $PP_j^kP$ is positive, $P P_j^k P|\psi\>=P_j^k|\psi\>\neq 0$,
and
$$(PP_j^kP)^2= P P_j^k P P_j^k P =P P_j^k P_j^k P=P P_j^k P.$$ (ii)
follows from $(PP_j^kP)(PP_{j'}^kP) = PP_j^k P_{j'}^kP =0$ when
$j\neq j'$, and (iii) from $\sum_{j=1}^L PP_j^kP=\sum_{j=1}^L P_j^kP
= P$. \hfill $\Box$

\section{Characterization of locally undetermined states}

Given a multipartite quantum state $\rho$ in Hilbert space $\ch$, we
can easily determine its reduced state of any proper subset
$S\subseteq \{1,\dots, n\}\define \n$ by tracing out all the systems
not included in $S$. The converse of this problem is, however, very
complicated. On one hand, given states $\rho_{S_1},\dots,
\rho_{S_k}$ for some proper subsets $S_1,\dots,S_k\subseteq \n$, it
is generally very difficult to determine if they are compatible in
the sense that there exists some global state $\rho$ in $\ch$ such
that the reduced state of $\rho$ for the systems in $S_i$ is exactly
$\rho_{S_i}$ \cite{Co72,Hi03,Br04,HZG04,HZG052}. On the other hand,
even if the existence of such $\rho$ is pre-assumed, how to
construct one, and furthermore, to characterize precisely all such
states is still open. In the current paper, we only focus on local
determinability of $pure$ states among other $pure$ states in the
same Hilbert space. Allowing the considered states to be mixed will
extremely increase the complexity of the problem. We leave this
general case for further investigation.

\begin{definition} Let $|\psi\>$ be a pure state in the composite Hilbert space
$\ch$ with $dim(\h_i)=d_i$, $1\leq i\leq n$. We say that $|\psi\>$
is locally undetermined (among pure states) if there exists a pure
state $|\phi\>\in \h_1\otimes \cdots\otimes \h_n$ such that
$|\<\phi|\psi\>|\neq 1$, and $|\phi\>$ shares the same $n$
$(n-1)$-party reduced states with $|\psi\>$, i.e.
\begin{equation}\label{eq-reducedstate}
\tr_i |\psi\>\<\psi|=\tr_i |\phi\>\<\phi|\ : \ i=1,\dots,n.
\end{equation}
\end{definition}

It is worth noting that once the $n$ $(n-1)$-party reduced states
$\tr_i |\psi\>\<\psi|$, $1\leq i\leq n$, are specified, all
$m$-party reduced states are specified as well for $m<n$. We denote
by $R(\psi)$ the set of $|\phi\>$ satisfying
Eq.(\ref{eq-reducedstate}). Then $|\psi\>$ is locally determined if
and only if $|R(\psi)|=1$ where for a set $A$, $|A|$ is the
cardinality of $A$.

\begin{theorem}\label{thm-main}
A pure state $|\psi\>\in \h_1\otimes\cdots\otimes \h_n$ is locally
undetermined if and only if it is GSD. Furthermore,
\begin{equation}\label{eq-set}
R(\psi)=\left\{\sum_{j=1}^L \exp (\i\theta_j)\bigotimes_{i=1}^n
P_j^i|\psi\> : \mbox{$P_j^i$ are Schmidt projectors of } |\psi\>,
\mbox{ and } 0\leq \theta_1,\dots, \theta_L < 2\pi \right\}.
\end{equation}
\end{theorem}

{\it Proof.} For the sufficiency part, we need only prove that any
state $|\phi\>$ in the set $R(\psi)$ defined in Eq.(\ref{eq-set})
shares the same set of $(n-1)$-party reduced states with $|\psi\>$.
Let $ |\phi\>= \sum_{j=1}^L \exp(\i\theta_j)\bigotimes_{i=1}^n
P_j^i|\psi\>$ where $0\leq \theta_1,\dots, \theta_L < 2\pi$. Then
for each $1\leq k\leq n$, we have
\begin{eqnarray}
\tr_k |\phi\>\<\phi| &=& \tr_k \left(\sum_{j,j'=1}^L
\exp(\i(\theta_j-\theta_{j'}))\bigotimes_{i=1}^n
P_j^i|\psi\>\<\psi|\bigotimes_{i=1}^n P_{j'}^i\right)\\
&=& \sum_{j,j'=1}^L \exp(\i(\theta_j-\theta_{j'}))\bigotimes_{i\neq
k} P_j^i \tr_k (P_{j}^kP_{j'}^k|\psi\>\<\psi|)\bigotimes_{i\neq k}
P_{j'}^i\\
&=& \sum_{j=1}^L \bigotimes_{i\neq k} P_j^i \tr_k
(P_{j}^kP_{j}^k|\psi\>\<\psi|)\bigotimes_{i\neq k}
P_{j}^i\\
&=&\tr_k \left(\sum_{j=1}^L \bigotimes_{i=1}^n
P_j^i|\psi\>\<\psi|\bigotimes_{i=1}^n P_{j}^i\right).
\label{eq-tmp1}
\end{eqnarray}
On the other hand, from Item 3 of Definition \ref{def-decom} we can
easily check that the reduced state $\tr_k |\psi\>\<\psi|$ is
exactly described by Eq.(\ref{eq-tmp1}). That completes the proof of
the sufficiency part.

Now we turn to the necessity part. The main proof technology is from
Ref.\cite{WL08}. Suppose $|R(\psi)|>1$. Let $|\phi\>\in R(\psi)$ but
$|\<\phi|\psi\>|\neq 1$. For any $1\leq k\leq n$, since $\tr_k
|\psi\>\<\psi| = \tr_k |\phi\>\<\phi|$, there exists unitary
transformation $U_k$ on $\h_k$ such that $|\phi\>=U_k |\psi\>$. Let
$$U_k=\sum_{i=1}^{d_k} \exp(\i \theta_i^k) |i\>_k\<i|$$
 be the spectrum decomposition of
$U_k$, where $0\leq \theta_i^k<2\pi$. Then $\{|i_1\>_1\dots |i_n\>_n
: i_k=1,\dots, d_k, 1\leq k\leq n\}$ constitute a complete
orthonormal basis of $\ch$. Write the decomposition of $|\psi\>$
under this basis as
\begin{equation}\label{eq:psi}
|\psi\>=\sum_{i_1=1}^{d_1}\dots\sum_{i_n=1}^{d_n} C_{i_1\dots i_n}
|i_1\>_1\dots|i_n\>_n=\sum_I C_I |I\>
\end{equation}
 where $I=i_1\dots i_n$. It is
easy to check that for any $1\leq  k\leq n$ and $I$,
$$(|i\>_k\<i|) |I\> = \delta_{i,i_k}|I\>.$$ Thus we have for any
$j\neq k$,
\begin{eqnarray*}
U_k^\dag \otimes U_j |\psi\> &=& \sum_I C_I \sum_{i,i'} \exp[\i
(\theta_i^j-\theta_{i'}^k)](|i\>_j\<i|\otimes |i'\>_k\<i'|) |I\>\\
&=&\sum_I C_I \exp[\i (\theta_{i_j}^j-\theta_{i_k}^k)]  |I\>,
\end{eqnarray*}
and then $C_I \exp[\i (\theta_{i_j}^j-\theta_{i_k}^k)]=C_I$ from the
fact that $U_k^\dag \otimes U_j |\psi\>=|\psi\>$. This means that
whenever $C_I\neq 0$ in the decomposition Eq.(\ref{eq:psi}) of
$|\psi\>$, we have $\theta_{i_j}^j=\theta_{i_k}^k$ for any $j\neq
k$.

Denote by $\ind$ the set of all the basis states $|I\>$ on which
$|\psi\>$ has nonzero component, i.e.,
$$\ind =\{ |I\> : C_I = \<I|\psi\> \neq 0 \}.$$
Then for any $|I\>, |J\>\in \ind$, if $I$ and $J$ are adjacent,
i.e., there exists $k_0$ such that $i_{k_0}=j_{k_0}$, then for any
$1\leq k\leq n$, we have
\begin{equation}\label{eq-theta}
\theta_{i_k}^k=\theta_{i_{k_0}}^{k_0}=\theta_{j_{k_0}}^{k_0}=\theta_{j_k}^k.
\end{equation}
Furthermore, if $I$ and $J$ are connected, i.e., there exists
$|I^1\>,\dots,|I^m\>\in \ind$ such that $I^1=I, I^m=J$ and for each
$1\leq i\leq m-1$, $I^i$ and $I^{i+1}$ are adjacent, then from
Eq.(\ref{eq-theta}) we have
\begin{equation}\label{eq-thetaeq}
\theta_{i_k}^k=\theta_{i^1_{k}}^{k}=\dots=\theta_{i^m_k}^k=\theta_{j_k}^k
\end{equation}
for any $1\leq k\leq n$.

Partition $\ind$ into subsets $\ind_1,\dots,\ind_L$ such that for
any $|I\>,|J\>\in \ind$, $|I\>$ and $|J\>$ belong to a same $\ind_l$
if and only if there are connected. According to this division,
$|\psi\>$ can be rewritten as
\begin{equation}
|\psi\>=\sum_{j=1}^L \sum_{|I\>\in \ind_j} C_I |I\>.
\end{equation}
We claim that $L>1$. Otherwise any $|I\>, |J\>$ in $\ind$ are
connected, hence from Eq.(\ref{eq-thetaeq}),
$$|\phi\>=U_k|\psi\>=\sum_{i=1}^{d_k}\sum_{|I\>\in \ind} C_I
[\exp(\i \theta_i^k) |i\>_k\<i|]|I\>=\sum_{|I\>\in \ind} C_I \exp(\i
\theta_{i_k}^k) |I\>=\exp(\i \theta_k)|\psi\>,$$ a contradiction.

Now for each $1\leq k\leq n$ and $1\leq j\leq L$, let
$$A_j^k = \{|i_k\>_k : \mbox{there exist $|i_j\>_j$, $1\leq j\leq n$ but $j\neq k$, such that } |i_1\>\dots|i_n\>\in \ind_j\},$$   and
$P_j^k = \sum_{|i_k\>\in A_j^k}|i_k\>\<i_k|$ be a projector on
$\h_k$. Take arbitrarily $|l_k\>\in A_j^k$ and $|l'_k\>\in A_{j'}^k$
for $j\neq j'$. By definition, there exist $|I\>\in \ind_j$ and
$|I'\>\in \ind_{j'}$ such that $i_k=l_k$ and $i_k'=l_k'$. Since $I$
and $I'$ are not adjacent (otherwise $j=j'$), we have $i_k\neq
i_k'$, and hence $\<l_k|l_k'\>=\<i_k|i_k'\>=0$. That is, the
projectors $P_j^k : j=1,\dots, L$ are pairwise orthogonal for any
fixed $k$.

If $|I\>\in \ind_j$, then $i_k\in A_j^k$ and hence $P_j^k|I\>=|I\>$.
If $|I\>\not\in \ind_j$, then by definition, for any $J\in \ind_j$,
$i_k\neq j_k$. So $i_k\not\in A_j^k$ and $P_j^k|I\>=0$. In a word,
for any $|I\>\in \ind$ and $1\leq k\leq n$,
$$P_j^k|I\>=\left\{
                                                 \begin{array}{ll}
                                                   |I\>, & \hbox{if } |I\>\in \ind_j,
                                                   \\ \\
                                                   0, & \hbox{if }  |I\>\not\in \ind_j.
                                                 \end{array}
                                               \right.
$$
We derive further that for any $j=1,\dots,n$,
\begin{eqnarray}\label{eq-temp2}
\bigotimes_{i=1}^n P_j^i|\psi\> &=&  \sum_{j'=1}^L
 \sum_{|I\>\in \ind_{j'}} C_I\bigotimes_{i=1}^n P_j^i|I\>=   \sum_{|I\>\in \ind_{j}} C_I|I\>,
\end{eqnarray}
and hence $|\psi\>$ is GSD.

Finally, let $R=\left\{\sum_{j=1}^L \exp
(\i\theta_j)\bigotimes_{i=1}^n P_j^i|\psi\> : 0\leq \theta_1,\dots,
\theta_L < 2\pi\right \}$. We need to show $R(\psi)=R$ to finish the
proof of this theorem. Note that at the sufficiency part, we have
already proved $R\subseteq R(\psi)$. To show the opposite side, let
$|\phi\>\in R(\psi)$ and fix arbitrarily $k$. Then
\begin{eqnarray*}
|\phi\>=U_k|\psi\>&=&\sum_{i=1}^{d_k}\sum_{j=1}^L \sum_{|I\>\in
\ind_j}
C_I [\exp(\i \theta_i^k) |i\>_k\<i|]|I\>\\
&=&\sum_{j=1}^L\sum_{|I\>\in \ind_j} C_I \exp(\i \theta_{i_k}^k)
|I\>\\
&=&\sum_{j=1}^L \exp(\i \theta_j)\sum_{|I\>\in \ind_j} C_I |I\>\\
&=&\sum_{j=1}^L \exp(\i \theta_j) \bigotimes_{i=1}^n P_j^i|\psi\>,
\end{eqnarray*}
where the fourth equation follows from Eq.(\ref{eq-thetaeq}) and the
last from Eq.(\ref{eq-temp2}). \hfill $\Box$

\vspace{1em} If we are not concerned with the set $R(\psi)$, a
simpler criteria for local determinability can be derived, as the
following corollary states.
\begin{corollary}
Pure state $|\psi\>\in \ch$ is locally undetermined if and only if
for each $i=1,\dots,n$, there exist projectors $P_1^i$ and $P_2^i$
satisfying $P_1^i|\psi\>\neq 0$, $P_2^i|\psi\>\neq 0$, and $P_1^i
\perp P_2^i$, such that $$|\psi\>= \bigotimes_{i=1}^n P_1^i |\psi\>
+ \bigotimes_{i=1}^n P_2^i|\psi\>.$$
\end{corollary}
{\it Proof.} The sufficiency part is direct from Theorem
\ref{thm-main}. For the necessity part, suppose $|\psi\>$ is locally
undetermined. Then from Theorem \ref{thm-main}, projectors $\{Q_j^i
: i=1,\dots,n; j=1,\dots, L\}$, $L\geq 2$, can be found such that
for any $i$, $\{Q_j^i : j=1,\dots, L\}$ are pairwise orthogonal in
$\h_i$, $Q_j^i|\psi\>\neq 0$, and
\begin{equation}
|\psi\>=\sum_{j=1}^L \bigotimes_{i=1}^n Q_j^i|\psi\>.
\end{equation}
Let $P_1^i=Q_1^i$ and $P_2^i=\sum_{j=2}^L Q_j^i$. Then $P_1^i\perp
P_2^i$, $P_1^i|\psi\>\neq 0$, and
\begin{eqnarray*}
\bigotimes_{i=1}^n P_2^i|\psi\>&=&\left[\sum_{j_1,\dots,j_n=2}^L
\bigotimes_{i=1}^n Q_{j_i}^i \right]\sum_{j=1}^L \bigotimes_{i=1}^n
Q_j^i
|\psi\>\\
&=&\sum_{j=1}^L\sum_{j_1,\dots,j_n=2}^L\delta_{j,j_1}\cdots\delta_{j,j_n}\bigotimes_{i=1}^nQ_j^i|\psi\>\\
&=&\sum_{j=2}^L \bigotimes_{i=1}^n Q_j^i|\psi\>.
\end{eqnarray*}
Hence we have
$$\bigotimes_{i=1}^n P_1^i |\psi\> + \bigotimes_{i=1}^n
P_2^i|\psi\>=\bigotimes_{i=1}^n Q_1^i |\psi\> + \sum_{j=2}^L
\bigotimes_{i=1}^n Q_j^i|\psi\>=|\psi\>.$$ Now we show that
$P_2^i|\psi\>\neq 0$ for any $i$. Otherwise $\bigotimes_{i=1}^n
P_2^i|\psi\>=0$, and then $|\psi\>=\bigotimes_{i=1}^n P_1^i
|\psi\>$. So we derive that $Q_j^i|\psi\>=0$ for any $j>1$, which is
a contradiction. \hfill $\Box$

\vspace{1em}

Theorem \ref{thm-main} provides a necessary and sufficient condition
for a pure multipartite state to be locally undetermined by means of
generalized Schmidt decomposability. The Schmidt projectors are,
however, hard to find in general. In the next theorem, by employing
(ordinary bipartite) Schmidt decomposition for some proper partition
of the original parties, we obtain a more practical method to
determine the local determinability of a pure state.

\begin{theorem}\label{thm:schdecom}
Let $|\psi\>$ be a pure state in $\ch$. If $|\psi\>$ is locally
undetermined, then for any $1\leq k\leq n$ there exists a (ordinary)
Schmidt decomposition
\begin{equation}\label{eq-tp-3-1}
|\psi\>=\sum_{i=1}^{M_k}\sqrt{\lambda_i}|i\>_k |\psi_i\>_{\bar{k}},
\hspace{2em} \lambda_1,\dots,\lambda_{M_k}>0
\end{equation}
of $|\psi\>$ when treated as a bipartite state between $\h_k$ and
$\h_{\bar{k}}$, and a complete partition $S_1,\dots,S_L$, $L\geq 2$,
of $\{1,\dots,M_k\}\define\m_k$ such that for any $j\neq k$, $1\leq
l\neq l'\leq L$, $r\in S_l$, $t\in S_{l'}$, it holds that
\begin{equation}\label{eq-tp-23-1}
\rho^{\psi_r}_j \perp \rho^{\psi_{t}}_j.
\end{equation} Furthermore
\begin{equation}
R(\psi)=\left\{\sum_{j=1}^L \exp(\i\theta_j)\sum_{i\in
S_j}\sqrt{\lambda_i}|i\>_k|\psi_i\>_{\bar{k}}: \mbox{$S_j$ satisfy
the conditions above}, \mbox{ and } 0\leq \theta_1,\dots, \theta_L <
2\pi\right \}.
\end{equation}

Conversely, if there exists $1\leq k\leq n$ such that a Schmidt
decomposition of $|\psi\>$ and a partition of $\m_k$ satisfying the
conditions presented above can be found, then $|\psi\>$ is locally
undetermined.

\end{theorem}
{\it Proof.} Suppose $|\psi\>$ is locally undetermined. Then from
Theorem \ref{thm-main} and Proposition \ref{prop-main}, there exist
$\{P_j^i : i=1,\dots, n; j=1,\dots, L\}$, $L>1$, such that for any
$i$, $P^i_j : j=1,\dots, L$ constitute a projective measurement in
$\supp(\rho_i^\psi)$, and $|\psi\>= \sum_{j=1}^L \bigotimes_{i=1}^n
P_j^i |\psi\>$.

For any $1\leq k\leq n$ and $1\leq j\leq L$, let
\begin{equation}\label{eq-tp-4-1}
\bigotimes_{i=1}^n P_j^i |\psi\>=\sum_{i\in
S_j}\sqrt{\lambda_i}|i\>_k|\psi_i\>_{\bar{k}},
\end{equation}
be a Schmidt decomposition of the unnormalized state
$\bigotimes_{i=1}^n P_j^i|\psi\>$ when treated as a bipartite state
between $\h_k$ and $\h_{\bar{k}}$, where for each $i\in S_j$,
$\lambda_i>0$, and $|i\>$ and $|\psi_i\>$ are normalized. It is easy
to check that $P_j^k|i\>_k = |i\>_k$ and
$P_j^{k'}|\psi_i\>_{\bar{k}}=|\psi_i\>_{\bar{k}}$ for any $k'\neq k$
and $i\in S_j$.

For any $j\neq j'$, $i\in S_j$, and $i'\in S_{j'}$, we have $
_k\<i|i'\>_k=\hspace{0em} _k\<i|P_j^k P_{j'}^k |i'\>_k=0$ since
$P_j^k\perp P_{j'}^k $, and $
_{\bar{k}}\<\psi_i|\psi_{i'}\>_{\bar{k}}=\hspace{0em}
_{\bar{k}}\<\psi_i|P_j^{k'} P_{j'}^{k'} |\psi_{i'}\>_{\bar{k}}=0$
since $P_j^{k'}\perp P_{j'}^{k'} $. As a consequence,
\begin{equation}\label{eq-tp-4-2}
|\psi\>=\sum_{j=1}^L \bigotimes_{i=1}^n P_j^i |\psi\>=\sum_{j=1}^L
\sum_{i\in S_j}\sqrt{\lambda_i}|i\>_k|\psi_i\>_{\bar{k}},
\end{equation}
is a Schmidt decomposition of $|\psi\>$.

For any $j\neq k$, $1\leq l\neq l'\leq L$, $r\in S_l$, and $t\in
S_{l'}$, we have
$$\rho_j^{\psi_r} = \tr_{\bar{j}} |\psi_r\>_{\bar{k}}\<\psi_r| = \tr_{\bar{j}}
\left[P_l^j|\psi_r\>_{\bar{k}}\<\psi_r|P_l^j\right]=P_l^j
\left[\tr_{\bar{j}}|\psi_r\>_{\bar{k}}\<\psi_r|\right]P_l^j$$ and

$$\rho_j^{\psi_t} = \tr_{\bar{j}}
|\psi_t\>_{\bar{k}}\<\psi_t| = \tr_{\bar{j}}
\left[P_{l'}^j|\psi_s\>_{\bar{k}}\<\psi_s|P_{l'}^j\right]=P_{l'}^j
\left[\tr_{\bar{j}}|\psi_s\>_{\bar{k}}\<\psi_s|\right]P_{l'}^j.$$ So
$\rho_j^{\psi_r}\perp \rho_j^{\psi_t}$ from the orthogonality of
$P_l^j$ and $P_{l'}^j$.

Furthermore, from Eqs.(\ref{eq-set}) and (\ref{eq-tp-4-1}) we derive
that
\begin{eqnarray*}
R(\psi)&=&\sum_{j=1}^L\exp(\i\theta_j) \bigotimes_{i=1}^n P_j^i|\psi\>\\
&=&\sum_{j=1}^L \exp(\i\theta_j)\sum_{i\in
S_j}\sqrt{\lambda_i}|i\>_k|\psi_i\>_{\bar{k}}.
\end{eqnarray*}

Conversely, suppose there exists $1\leq k\leq n$ such that a Schmidt
decomposition of $|\psi\>$ and a partition of $\m_k$ satisfying the
conditions presented in the Theorem can be found. For any $j\neq k$
and $1\leq l\leq L$, let $\h_{l}^j=\supp\{\rho^{\psi_i}_j : i\in
S_l\}$, and $P_l^j$ be the projector onto $\h_l^j$. Let $P_l^k
=\sum_{i\in S_l} |i\>_k\<i|$. Then it is obvious that for any $1\leq
j\leq n$, $P_l^j$ are pairwise orthogonal projectors on $\h_j$, and
$P_l^j|\psi\>\neq 0$. Furthermore, for any $j\neq k$, $1\leq i \leq
M_k$, $1\leq l\leq L$, we have $P_l^k|i\>_k = \delta_{i\in S_l}
|i\>_k$ $P_l^j|\psi_i\>_{\bar{k}}=\delta_{i\in S_l}
|\psi_i\>_{\bar{k}}$, where $\delta_{i\in A}$ equals 1 if $i\in A$
while 0 if $i\not\in A$. Hence we deduce that
\begin{eqnarray*}
\sum_{l=1}^L \bigotimes_{j=1}^n P_l^j |\psi\> &=& \sum_{l=1}^L
\bigotimes_{j=1}^n P_l^j \left[\sum_{i=1}^{M_k}\sqrt{\lambda_i}|i\>_k |\psi_i\>_{\bar{k}}\right]\\
&=& \sum_{l=1}^L \sum_{i=1}^{M_k}\sqrt{\lambda_i}\left[P_l^k
|i\>_k\right]\left[ \bigotimes_{j\neq k} P_l^j|\psi_i\>_{\bar{k}}
\right] \\
&=& \sum_{l=1}^L \sum_{i\in S_l}\sqrt{\lambda_i}|i\>_k
|\psi_i\>_{\bar{k}} \\
&=& |\psi\>.
\end{eqnarray*}
Then $|\psi\>$ is locally undetermined from Theorem
\ref{thm-main}.\hfill $\Box$

\vspace{1em}

Following Theorem \ref{thm:schdecom}, we can obtain a simple way to
check whether $|\psi\>$ is locally undetermined when one of the
1-party reduced states has distinct nonzero eigenvalues.

\begin{corollary}\label{thm-neqeigen}
Suppose $|\psi\>\in \ch$ and there exists $1\leq k\leq n$ such that
$\rho^{\psi}_k$ has distinct nonzero eigenvalues, and suppose the
Schmidt decomposition of $|\psi\>$ when treated as a bipartite state
between $\h_k$ and $\h_{\bar{k}}$ has the form
\begin{equation}\label{eq-schdecom}
|\psi\>=\sum_{i=1}^{M_k}\sqrt{\lambda_i}|i\>_k |\psi_i\>_{\bar{k}}
\end{equation}
where $\lambda_1>\dots >\lambda_{M_k}$. Then $|\psi\>$ is locally
undetermined if and only if there exists a complete partition
$S_1,\dots,S_L$ of $\{1,\dots,M_k\}\define\m_k$
 such that for any $j\neq k$, $1\leq l\neq l'\leq L$, $r\in S_l$, $t\in S_{l'}$, it
holds that
\begin{equation}\label{eq-tp-23-1-2}
\rho^{\psi_r}_j \perp \rho^{\psi_{t}}_j.
\end{equation}
 Furthermore
\begin{equation}
R(\psi)=\left\{\sum_{j=1}^L \exp(\i\theta_j)\sum_{i\in
S_j}\sqrt{\lambda_i}|i\>_k|\psi_i\>_{\bar{k}}: \mbox{$S_j$ satisfy
the conditions above}, \mbox{ and } 0\leq \theta_1,\dots, \theta_L <
2\pi\right \}.
\end{equation}

Particularly, if $M_k=2$, then $|\psi\>$ is locally undetermined if
and only if for any $j\neq k$,
$$\rho^{\psi_1}_j  \perp \rho^{\psi_2}_j,$$ and
\begin{equation}
R(\psi)=\left\{\sqrt{\lambda_1}|1\>_k|\psi_1\>_{\bar{k}}+\exp(\i\theta)\sqrt{\lambda_2}|2\>_k|\psi_2\>_{\bar{k}}
: 0\leq \theta<2\pi \right \}.
\end{equation}
\end{corollary}
{\it Proof.} Notice that when $\rho^{\psi}_k$ has distinct nonzero
eigenvalues, the Schmidt decomposition of $|\psi\>$ under the
partition $\{k, \bar{k}\}$ of $\{1,\dots, n\}$ has a unique form as
in Eq.(\ref{eq-schdecom}). Then the corollary follows directly from
Theorem \ref{thm:schdecom}. \hfill $\Box$

\vspace{1em}

\begin{theorem}\label{thm-gschdec}
Suppose $|\psi\>\in \ch$ for $n\geq 3$ and
\begin{equation}\label{eq-schde}
|\psi\>=\sum_{i=1}^m \sqrt{\lambda_i} |i\>_1\dots|i\>_n
\end{equation}
is completely GSD where $m\leq \min\{d_k : 1\leq k\leq n\}$,
$\lambda_1,\dots,\lambda_m>0$, and $\{|i\>_k : i=1,\dots,d_k\}$ is
an orthonormal basis for each $\h_k$. Then $|\psi\>$ is locally
undetermined if and only if $m>1$, and when $m>1$,
$$R(\psi)=\left\{\sum_{i=1}^m \sqrt{\lambda_i}\exp(\i
\theta_i)|i\>_1\dots|i\>_n : 0\leq \theta_1,\dots,\theta_m<
2\pi\right\}.$$
\end{theorem}
{\it Proof.} First it is easy to check that $|\psi\>$ is locally
undetermined if and only if $m>1$. Suppose $m>1$. Then from Theorem
\ref{thm-main}, any $|\phi\>\in R(\psi)$ has the form
$|\phi\>=\sum_{j=1}^L \exp(\i\theta_j)\bigotimes_{i=1}^n
P_j^i|\psi\>$ where $L>1$, $0\leq \theta_1,\dots, \theta_L \leq
2\pi$, $P_j^i : j=1,\dots, L$ are pairwise orthogonal projectors on
$\h_i$, and
\begin{equation}\label{eq-tp-7-9-1}
|\psi\>=\sum_{j=1}^L \bigotimes_{i=1}^n P_j^i|\psi\>.
\end{equation}

Fix arbitrarily $1\leq j\leq L$. For any $1\leq i, i'\leq n$, we
observe that
$$P_j^i|\psi\>=P_j^{i'}|\psi\>=\bigotimes_{i=1}^n P_j^i|\psi\>,$$
hence from Eq.(\ref{eq-schde})
\begin{equation}\label{eq-tp22-1}
\sum_{l=1}^m \sqrt{\lambda_l}|l\>_1\dots [P_j^i|l\>_i]\dots |l\>_n=
\sum_{l=1}^m \sqrt{\lambda_l}|l\>_1\dots [P_j^{i'}|l\>_{i'}]\dots
|l\>_n,
\end{equation}
and $_i\<l|P_j^i|l\>_i=\mbox{}_{i'}\<l|P_j^{i'}|l\>_{i'}$ by
multiplying both sides by $_1\<l|\dots\ _n\<l|$. That is, the
quantity $_i\<l|P_j^i|l\>_i$ is independent of $i$. Let
$\alpha_{j,l}=\mbox{} _i\<l|P_j^i|l\>_i\geq 0$. Then from
Eq.(\ref{eq-tp22-1}) we have
\begin{equation}\label{eq-tp22-2}
P_j^i|l\>_i = \alpha_{j,l}|l\>_i.
\end{equation}
Furthermore, from the relation $$\sum_{j=1}^L
P_j^i|\psi\>=\sum_{j=1}^L \bigotimes_{i=1}^n P_j^i|\psi\>=|\psi\>$$
we can deduce that $\sum_{j=1}^L \alpha_{j,l}=1$ for each $1\leq
l\leq m$. On the other hand, taking Eq.(\ref{eq-tp22-2}) back into
Eq.(\ref{eq-tp-7-9-1}) we have
$$\sum_{l=1}^m \sqrt{\lambda_l}|l\>_1\dots  |l\>_n=
\sum_{l=1}^m \sum_{j=1}^L\sqrt{\lambda_l} \bigotimes_{i=1}^n
\left(P_j^{i}|l\>_{i}\right)= \sum_{l=1}^m
\sum_{j=1}^L\sqrt{\lambda_l} \alpha_{j,l}^n|l\>_1\dots  |l\>_n.$$ So
$\sum_{j=1}^L \alpha_{j,l}^n=1$, and hence for each $1\leq l\leq m$,
there exists one and only one $j$, denoted by $j_l$ such that
$\alpha_{j,l}=1$; other $\alpha_{j,l}$ equal 0. Now we can calculate
that
\begin{eqnarray*}
|\phi\>&=&\sum_{j=1}^L \exp(\i\theta_j)\bigotimes_{i=1}^n
P_j^i|\psi\>\\
&=&\sum_{j=1}^L \exp(\i\theta_j)\sum_{l=1}^m
\sqrt{\lambda_l}\alpha_{j,l}^n|l\>_1\dots  |l\>_n\\
&=&\sum_{l=1}^m \sqrt{\lambda_l}\exp(\i\theta_{j_l})|l\>_1\dots
|l\>_n.
\end{eqnarray*}
That completes the proof of the theorem. \hfill $\Box$

\begin{corollary}
Suppose $|\psi\>$ is a pure state in $n$-qubit system, i.e.,
$\dim(\h_i)=2$ for each $1\leq i\leq n$. Then $|\psi\>$ is locally
undetermined if and only if $|\psi\>$ is completely GSD (or, as
stated in \cite{WL08}, $|\psi\>$ is a generalized GHZ state):
$|\psi\>= \alpha |0\>_1\dots |0\>_n + \beta |1\>_1\dots |1\>_n$ with
$\alpha
>0$ and $\beta>0$. Furthermore, if $|\psi\>$ is locally undetermined, then
$$R(\psi)=\{\mbox{all maximally entangled states in
$\mathbb{C}^2\otimes \mathbb{C}^2$ space}\}$$ when $n=2$ and $\alpha
= \beta$; otherwise
\begin{equation}\label{eq-tp-24-1}
R(\psi)=\{\ \alpha |0\>_1\dots |0\>_n + \exp(\i\theta) \beta
|1\>_1\dots |1\>_n : 0\leq \theta < 2\pi\}.
\end{equation}
\end{corollary}
{\it Proof.} From Theorem \ref{thm-main}, $|\psi\>$ is locally
undetermined if and only if there exists an orthonormal basis,
denoted by $\{|\widehat{0}\>_i, |\widehat{1}\>_i\}$, for each $\h_i$
such that
\begin{eqnarray}
|\psi\>&=&\left(\bigotimes_{i=1}^n
|\widehat{0}\>_i\<\widehat{0}|\right)|\psi\>
+\left(\bigotimes_{i=1}^n |\widehat{1}\>_i\<\widehat{1}|\right)|\psi\>\\
&=&\widehat{\alpha} |\widehat{0}\>_1\dots|\widehat{0}\>_n +
\widehat{\beta}
|\widehat{1}\>_1\dots|\widehat{1}\>_n,\label{eq-special}
\end{eqnarray}
where
$\widehat{\alpha}=\hspace{0em}_1\<\widehat{0}|\dots\hspace{0em}_n\<\widehat{0}|\psi\>$
and
$\widehat{\beta}=\hspace{0em}_1\<\widehat{1}|\dots\hspace{0em}_n\<\widehat{1}|\psi\>$.
From the fact that $|\widehat{0}\>_i\<\widehat{0}|\psi\>\neq 0$ for
each $i$, we know $\widehat{\alpha}\neq 0$. Similarly, it holds that
$\widehat{\beta}\neq 0$. Let $\widehat{\alpha}=\alpha
\exp(\i\theta_\alpha)$ and $\widehat{\beta}=\beta
\exp(\i\theta_\beta)$ where $\alpha=|\widehat{\alpha}|>0$ and
$\beta=|\widehat{\beta}|>0$. Then we have $|\psi\>= \alpha
|0\>_1\dots |0\>_n + \beta |1\>_1\dots |1\>_n$ by, say, letting
$|0\>_1=\exp(\i\theta_\alpha)|\widehat{0}\>_1$,
$|1\>_1=\exp(\i\theta_\beta)|\widehat{1}\>_1$, and
$|0\>_i=|\widehat{0}\>_i$ and $|1\>_i=|\widehat{1}\>_i$ for $i\geq
2$.

When $n=2$ and $\alpha=\beta$, we have
$\rho_1^{\psi}=\rho_2^{\psi}=I/2$. Hence $|\phi\>\in R(\psi)$ if and
only if $|\phi\>$ is a maximally entangled states in
$\mathbb{C}^2\otimes \mathbb{C}^2$. Furthermore, we can show that
$R(\psi)$ has the form in Eq.(\ref{eq-tp-24-1}) by Corollary
\ref{thm-neqeigen} for the case of $n=2$ and $\alpha \neq \beta$
while by Theorem \ref{thm-gschdec} for the case of $n\geq 3$. \hfill
$\Box$

\vspace{1em}

To conclude this section, we would like to point out that the
techniques developed in this section can be used in locally
determining an $n$-party pure state when only a proper subset of the
$(n-1)$-party reduced states are specified. To be specific, we call
a pure state $|\psi\>\in \ch$ $S$-locally undetermined for some
$S\subseteq \{1,\dots,n\}$ and $|S|>1$ if there exists a pure state
$|\phi\>\in \h_1\otimes \cdots\otimes \h_n$ such that
$|\<\phi|\psi\>|\neq 1$, and for each $k\in S$, $|\phi\>$ shares the
same $(n-1)$-party reduced states with $|\psi\>$ when tracing out
the $k$th subsystem, i.e., $\tr_k |\psi\>\<\psi| = \tr_k
|\phi\>\<\phi|$. $R_S(\psi)$ can be defined similarly. Then all the
results presented in this section can be extended to this general
notion of $S$-local determinability by simply replacing the index
range $\{1,\dots, n\}$ by $S$. For example, the result corresponding
to Theorem \ref{thm-main} can be stated as follows: $|\psi\>$ is
$S$-locally undetermined if and only if there exist projectors
$\{P_j^i : i\in S; j=1,\dots, L\}$, $L> 1$, such that for any fixed
$i\in S$, $P_j^i : j=1,\dots, L$ are pairwise orthogonal projectors
on $\h_i$, $P_j^i|\psi\>\neq 0$, and $|\psi\>=\sum_{j=1}^L
\bigotimes_{i\in S} P_j^i|\psi\>.$ Furthermore, when $|\psi\>$ is
$S$-locally undetermined, then
$$R_S(\psi)=\left\{\sum_{j=1}^L \exp (\i\theta_j)\bigotimes_{i\in S}
P_j^i|\psi\> : \mbox{$P_j^i$ satisfy the conditions above}, \mbox{
and } 0\leq \theta_1,\dots, \theta_L < 2\pi \right\}.$$

\section{Application in distributed consensus}

The purpose of this section is, similar to that of \cite{DP05}, to
characterize the exact quantum resource that is sufficient and
necessary to solve distributed consensus problem, by applying the
notion of local determinability. As pointed out in Introduction,
D'Hondt and Panangaden considered anonymous network setting in which
all agents are completely identical without an individual name to
distinguish them. As a result, the protocols executed by all agents
are the same, and the shared entangled states, as a quantum resource
to solve the problem, is invariant under any permutation of agent
subspaces. Here in the current paper, however, we relax this
constraint to consider more general network setting which is not
necessarily anonymous. Interestingly, we find that locally
undetermined pure states play a key role in solving distributed
consensus for this general network, just like GHZ-like states play
in anonymous setting.

\begin{theorem}
Suppose a set of physically separated agents $A_1,\dots,A_n$ share a
multipartite pure quantum state $|\psi\>\in \ch$ where agent $A_i$
holds the particle in $\h_i$. Furthermore, communication between
them, classical or quantum, is unreliable and local ancillary
quantum systems are forbidden. Then there exists a totally correct
protocol for these agents to reach a consensus if and only if
$|\psi\>$ is locally undetermined (equivalently, $|\psi\>$ is GSD).
\end{theorem}
{\it Proof.} The sufficiency part is easy from Theorem
\ref{thm-main} and Proposition \ref{prop-main}. Suppose $|\psi\>$ is
locally undetermined. Then there exist projectors $\{P_j^i :
i=1,\dots,n; j=1,\dots, L\}$, $L\geq 2$, such that for any fixed
$i$, $P_j^i : j=1,\dots, L$ constitute a projective measurement in
$\supp(\rho_i^\psi)$, and $|\psi\>=\sum_{j=1}^L \bigotimes_{i=1}^n
P_j^i|\psi\>$. Let $P_i$ be the projector to the ortho-complement of
$\supp(\rho_i^\psi)$ in $\h_i$. Then a simple but totally correct
protocol for these $n$ agents to reach a consensus is as follows:
agent $i$ performs the projective measurement $\{P_i,P_j^i :
j=1,\dots, L\}$ on his/her shared particle, and treat the
measurement outcome as the agreement they meet. Since the
probability of obtaining the outcome corresponding to $P_i$ is 0,
and for any $1\leq j_1,\dots,j_n\leq L$,
$$\bigotimes_{i=1}^n P_{j_i}^i |\psi\>= \bigotimes_{i=1}^n P_{j_i}^i \sum_{j=1}^L
\bigotimes_{i=1}^n P_j^i|\psi\>=\sum_{j=1}^L \bigotimes_{i=1}^n
P_{j_i}^iP_j^i|\psi\>,$$ we deduce that $\bigotimes_{i=1}^n
P_{j_i}^i |\psi\>\neq 0$ if and only if $j_1=\dots=j_n$. That is,
the agents will definitely get a common measurement outcome, and so
reach a consensus.

For the necessity part, we note that since communication between the
agents are unreliable, no classical post-processing is allowed for
the protocol to be totally correct. Furthermore, by assumption local
ancillary systems in their labs are also forbidden. As a
consequence, the only way for them to reach agreement is each
performing a projective measurement $\{Q_j^i : \sum_j Q_j^i =
I_{\h_i}\}$ and announcing the outcome as their consensus. Deleting
all the projectors $Q_j^i$ which satisfy $Q_j^i|\psi\>=0$ from
$\{Q_j^i\}$ we get a set of pairwise orthogonal projectors $P_j^i :
j=1,\dots, L; L\geq 2$ such that $\sum_{j=1}^L P_j^i \geq
I_{\supp(\rho^\psi_i)}$. Then for any $1\leq j,j'\leq L$ and $i\neq
i'$, $P_j^i\otimes P_{j'}^{i'}|\psi\> = \delta_{j,j'} P_j^i\otimes
P_{j}^{i'}|\psi\>$. So we have
$$P_{j}^{1}|\psi\>=\sum_{j'=1}^L P_{j'}^2\otimes
P_{j}^{1}|\psi\>=  P_{j}^2\otimes P_{j}^{1}|\psi\>= \dots =
\bigotimes_{i=1}^n P_j^i|\psi\>,$$ and then $$|\psi\>=\sum_{j=1}^L
P_{j}^{1}|\psi\>=\sum_{j=1}^L  \bigotimes_{i=1}^n P_j^i|\psi\>.$$
From Theorem \ref{thm-main}, $|\psi\>$ is locally undetermined.
\hfill $\Box$

\section{Conclusion}

In this paper, we investigate the problem of locally determining
multipartite pure states. Necessary and sufficient conditions under
which a pure state is locally undetermined among pure states, as
well as the precise form of all the pure states sharing the same set
of reduced states with it, are presented. As an application, we
prove that a locally undetermined pure state can serve as a quantum
resource to solve distributed consensus problem in a general network
setting. More importantly, such states are the only possible pure
states which can achieve this goal in a totally correct and
completely fault-tolerant way.

What concerns us in this paper is local determinability of pure
state among pure states. There are two natural extensions of this
issue: (i) to determine a pure state among all states, pure or
mixed; (ii) to determine a mixed state among all states. In fact,
Linden $et$ $al$.'s work \cite{LPW02,LW02} is in the framework of
(i). New techniques must be proposed to give solutions for these two
general problems. Furthermore, to explore properties of multipartite
pure entanglement by using the results and techniques developed in
this paper is also a direction worthwhile for further study.

\section*{Acknowledgement}

The authors thank the colleagues in the Quantum Computation and
Quantum Information Research Group for useful discussion. This work
was partially supported by the FANEDD under Grant No.~200755, the
863 Project under Grant No.~2006AA01Z102, and the Natural Science
Foundation of China (Grant Nos.~60503001, 60621062).

\bibliographystyle{unsrt}
\bibliography{part}
\end{document}